\documentclass[aps,pra,preprintnumbers,showpacs,tightenlines]{revtex4}

\usepackage{amssymb}
\usepackage{amsmath}
\usepackage{graphicx}
\usepackage{epsfig}
\usepackage{subfigure}
\usepackage{amsfonts}
\usepackage{CJK}

\begin{document}

\title{Fast quantum gate with superconducting flux qubits
coupled to a cavity}

\author{Chui-Ping Yang$^{1,2}$}
\address{$^1$Department of Physics, Hangzhou Normal
University, Hangzhou, Zhejiang 310036, China}

\address{$^2$State Key Laboratory of Precision Spectroscopy, Department of Physics,
East China Normal University, Shanghai 200062, China}

\date{\today}

\begin{abstract}
We present a way for fast implementation of a two-qubit controlled
phase gate with superconducting flux qubits coupled to a cavity. A
distinct feature of this proposal is that since only qubit-cavity
resonant interaction and qubit-pulse resonant interaction are
used, the gate can be performed much faster when compared with the
previous proposals. This proposal does not require adjustment of
the qubit level spacings during the gate operation. In addition,
neither uniformity in the qubit parameters nor exact placement of
qubits in the cavity is needed by this proposal.

\end{abstract}

\maketitle
\date{\today}

\begin{center}
\textbf{I. INTRODUCTION}
\end{center}

The physical system composed of circuit cavities and
superconducting qubits such as charge, phase and flux qubits has been
considered as one of the most promising candidates for quantum information
processing. This is because: (i) superconducting qubits and microwave
cavities can be fabricated with modern integrated circuit technology, (ii) a
superconducting qubit has relatively long decoherence time [1,2], and (iii)
a superconducting microwave cavity or resonator acts as a ``quantum bus''
which can mediate long-range and fast interaction between distant
superconducting qubits [3-5]. In addition, the strong coupling between the
cavity field and superconducting qubits, which is difficult to achieve with
atoms in a microwave cavity, was earlier predicted by theory [6,7] and has
been experimentally demonstrated [8,9]. All of these features make
superconducting qubit cavity QED very attractive for quantum information
processing.

Over the past few years, there is much experimental progress in quantum
information processing with superconducting qubits. Two-qubit
controlled-phase, controlled-NOT, $i$SWAP gates, or other two-qubit
entangling gates have been experimentally demonstrated with supercoducting
charge qubits coupled by a capacitor [10], phase qubits coupled via a
capacitor [11], and flux qubits coupled through mutual
inductance [12]. Also, three-qubit entangled states have been recently generated in
experiments by using superconducting phase qubits coupled via a capacitor [13]
or a superconducting phase qubit coupled to two microscopic two-level
systems [14]. On the other hand, based on cavity QED technique, two-qubit
quantum gates [4,15], two-qubit entanglement [5], two-qubit quantum
algorithm [5], and quantum information transfer [4] have been experimentally
demonstrated with superconducting charge qubits or transmon qubits coupled
to a cavity or resonator. Moreover, based on cavity QED, experimental
demonstration of three-qubit Toffoli gates [16-18], three-qubit entanglement
[19] and three-qubit quantum error correction [17] with superconducting
transmon qubits or phase qubits coupled to a resonator has been reported
recently. However, to the best of our knowledge, no experimental
demonstration of one of them with superconducting flux qubits in cavity QED
has been reported.

It is known that two-qubit controlled-phase (CP) gates plus single-qubit
gates form the building blocks of quantum information
processors [20]. Theoretical methods for implementing a two-qubit
CP gate [3,21-27] have been
presented with flux qubits (e.g., SQUID qubits) or charge-flux qubits based
on cavity QED technique. However, these methods have some disadvantages. For
instances: (i) the methods presented in [3,21] require adjustment of the
qubit level spacings during the operation; (ii) the methods proposed in [22,23]
require slowly changing the Rabi frequencies to satisfy the
adiabatic passage; and (iii) the approaches introduced in [24-27] require a
second-order detuning to achieve an off-resonant Raman coupling between two
relevant levels. Note that the adjustment of the qubit level spacings during
the gate operation is undesirable and also may cause extra decoherence. In
addition, when the adiabatic passage or a second-order detuning is applied,
the gate becomes slow (the operation time required for the gate
implementation is on the order of one microsecond to a few microseconds [23-25]).

In this paper, we propose an alternative method for realizing a two-qubit CP
gate with four-level superconducting flux qubits coupled to a cavity or
resonator. This proposal has the following advantages: (i) since only
qubit-cavity resonant interaction and qubit-pulse resonant interaction are
applied, the gate operation can be performed faster by two orders of
magnitude, when compared with the previous proposals [22-27] requiring a
second-order large detuning or adiabatic passage; (ii) the method does not
require adjustment of the qubit level spacings (which however was needed by
the previous proposals [3,21], thus decoherence caused by tuning the qubit
level spacings is avoided; and (iii) the qubits are not required to have
identical level spacings, therefore superconducting devices, which often
have considerable parameter nonuniformity, can be used in this proposal.
This work is interesting because it avoids most of the problems existing in
the previous proposals [3,21-27] and the gate speed is significantly
improved.

\begin{center}
\textbf{II. BASIC THEORY}
\end{center}

The flux qubits throughout this paper have four
levels $\left| 0\right\rangle ,$ $\left| 1\right\rangle ,$ $\left|
2\right\rangle ,$ and $\left| 3\right\rangle $ as depicted in Fig. 1. In
general, there exists the transition between the two lowest levels $\left|
0\right\rangle $ and $\left| 1\right\rangle $ [28], which however can be
made to be weak via increasing the potential barrier between the two levels $%
\left| 0\right\rangle $ and $\left| 1\right\rangle $ [1,29,30]. The qubits
with this four-level structure could be a radio-frequency superconducting
quantum interference device (rf SQUID) consisting of one Josephson junction
enclosed by a superconducting loop, or a superconducting device with three
Josephson junctions enclosed by a superconducting loop. For flux qubits, the
two logic states of a qubit are represented by the two lowest levels $\left|
0\right\rangle $ and $\left| 1\right\rangle .$

\begin{figure}[tbp]
\centerline{\includegraphics[bb=82 362 265 539, width=4.6 cm, clip]{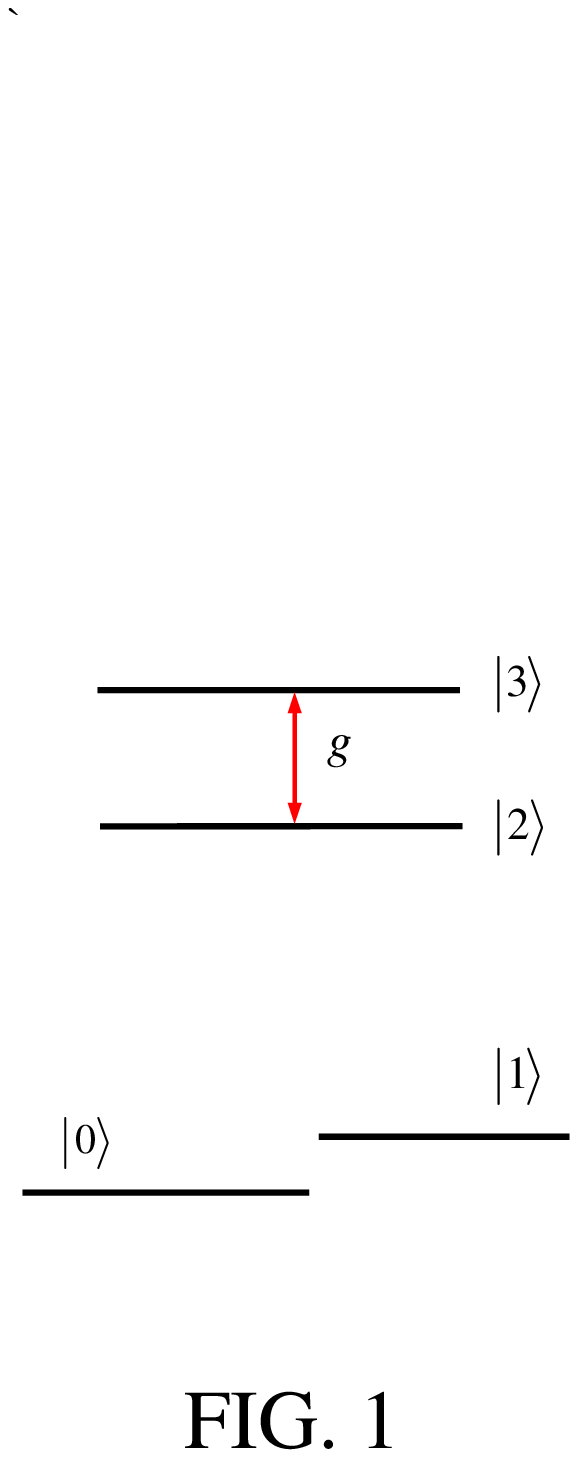}} %
\vspace*{-0.08in}
\caption{(Color online) Level diagram of a four-level flux qubit
(not drawn to scale), with weak transition between the two lowest levels.
The cavity mode is resonant with the transition between the top two levels.
$g$ is the coupling constant between the cavity mode and the
$\left| 2\right\rangle \leftrightarrow \left| 3\right\rangle $ transition.}
\label{fig:1}
\end{figure}

\subsection{Qubit-cavity resonant interaction}
Consider a flux qubit with
four levels as shown in Fig.~1. Suppose that the transition between the two
levels $\left| 2\right\rangle $ and $\left| 3\right\rangle $ is resonant
with the cavity mode. In the interaction picture and under the rotating-wave
approximation, the interaction Hamiltonian of the qubit and the cavity mode
is given by
\begin{equation}
H=\hbar g(a^{+}\sigma _{23}^{-}+\text{H.c.}),
\end{equation}
where $a^{+}$ and $a$ are the photon creation and annihilation operators of
the cavity mode, $g$ is the coupling constant between the cavity mode and
the $\left| 2\right\rangle \leftrightarrow \left| 3\right\rangle $
transition of the qubit, and $\sigma _{23}^{-}=\left| 2\right\rangle
\left\langle 3\right| $.

Based on the Hamiltonian (1), it can be easily found that the initial states
$\left| 3\right\rangle \left| 0\right\rangle _c$ and $\left| 2\right\rangle
\left| 1\right\rangle _c$ of the qubit and the cavity mode evolve as follows
\begin{eqnarray}
\left| 3\right\rangle \left| 0\right\rangle _c &\rightarrow &-i\sin \left(
gt\right) \left| 2\right\rangle \left| 1\right\rangle _c+\cos (gt)\left|
3\right\rangle \left| 0\right\rangle _c,  \nonumber \\
\left| 2\right\rangle \left| 1\right\rangle _c &\rightarrow &\cos \left(
gt\right) \left| 2\right\rangle \left| 1\right\rangle _c-i\sin \left(
gt\right) \left| 3\right\rangle \left| 0\right\rangle _c.
\end{eqnarray}
However, the state $\left| 0\right\rangle \left| 0\right\rangle _c$ remains
unchanged under the Hamiltonian (1).

The coupling strength $g$ may vary with different qubits due to non-uniform
device parameters and/or non-exact placement of qubits in the cavity.
Therefore, in the operation below, $g$ will be replaced by $g_1$ and $g_2$
for qubits $1$ and $2$, respectively.

\subsection{Qubit-pulse resonant interaction}

Consider a flux qubit with
four levels as depicted in Fig.~1, driven by a classical microwave pulse.
Suppose that the pulse is resonant with the transition between the two
levels $\left| i\right\rangle $ and $\left| j\right\rangle $ of the qubit.
Here, the level $\left| i\right\rangle $ is the lower energy level. In the
interaction picture and under the rotating-wave approximation, the
interaction Hamiltonian is given by
\begin{equation}
H_I=\hbar \left( \Omega _{ij}e^{i\phi }\left| i\right\rangle \left\langle
j\right| +\text{H.c.}\right) ,
\end{equation}
where $\Omega _{ij}$ is the Rabi frequency of the pulse and $\phi $ are the
initial phase of the pulse. Based on the Hamiltonian (3), it is
straightforward to show that a pulse of duration $t$ results in the
following state transformation
\begin{eqnarray}
\left| i\right\rangle &\rightarrow &\cos \Omega _{ij}t\left| i\right\rangle
-ie^{-i\phi }\sin \Omega _{ij}t\left| j\right\rangle ,  \nonumber \\
\left| j\right\rangle &\rightarrow &\cos \Omega _{ij}t\left| j\right\rangle
-ie^{i\phi }\sin \Omega _{ij}t\left| i\right\rangle ,
\end{eqnarray}
which can be completed within a very short time, by increasing the pulse
Rabi frequency $\Omega _{ij}$ (i.e., by increasing the intensity of the
pulse).

\begin{center}
\textbf{III. REALIZING A TWO-QUBIT CP GATE}
\end{center}

Let us consider two flux qubits 1
and 2. Each qubit has a four-level configuration as depicted in Fig.~1. To
begin with, it should be mentioned that during the gate implementation, the
following conditions are required, which are: (i) the cavity mode is
resonant with the $\left| 2\right\rangle \leftrightarrow \left|
3\right\rangle $ transition of each qubit, (ii) the cavity mode is highly
detuned (decoupled) from the transition between any other two levels, and
(iii) the pulse is resonant with the transition between two relevant levels
of each qubit but highly detuned (decoupled) from the transition between any
two irrelevant levels of each qubit.

For superconducting qubits, it is experimentally challenging to design the
qubits with identical level spacings, due to nonuniformity of the device
parameters. However, once superconducting qubits are designed, their level
spacings can be readily adjusted by changing the external parameters (e.g.,
changing the external magnetic flux for supercondcuting charge qubits, the
flux bias or current bias in the case of superconducting phase qubits and
flux qubits) [1,29-31]. Experimentally, it is difficult to adjust all
level spacings for two superconducting qubits to be identical, but it is
easy to adjust the level spacing between certain two levels to be the same
for the two qubits [32]. For instance, for the two flux qubits 1 and 2 here,
it is hard to make all of the $\left| 0\right\rangle \leftrightarrow \left| 1\right\rangle $
, $\left| 0\right\rangle \leftrightarrow \left| 2\right\rangle $, ..., and
$\left| 2\right\rangle \leftrightarrow \left|
3\right\rangle $ level spacings of qubit 1 to be, respectively, the same as
the $\left| 0\right\rangle \leftrightarrow \left| 1\right\rangle $,
$\left| 0\right\rangle \leftrightarrow \left| 2\right\rangle $, ..., and
$\left| 2\right\rangle \leftrightarrow \left|
3\right\rangle $ level spacings of qubit 2, by adjusting the
level spacings of the two qubits. But, the level spacing between certain two
levels (e.g., the levels $\left|
2\right\rangle $ and $\left| 3\right\rangle $) for the two qubits 1 and 2 can be
adjusted to be identical, which can be achieved by changing the external flux biases
applied to the superconducting loops of qubits 1 and 2. Thus, the first condition described
above can be achieved since one can set the level spacing between the two
levels $\left| 2\right\rangle $ and $\left| 3\right\rangle $ to be
the same for qubits 1 and 2, as discussed here. In addition, the second
and third conditions above can be also achieved via prior adjustment of the
qubit level spacings before the operation. Having these in mind, we now give
a detailed discussion on implementing a two-qubit CP gate.

For two qubits, there are a total number of four computational basis states,
denoted by $\left| 00\right\rangle ,\left| 01\right\rangle ,\left|
10\right\rangle ,$ and $\left| 11\right\rangle $, respectively. A two-qubit
CP gate is described by
\begin{equation}
\left| \epsilon _1\epsilon _2\right\rangle \rightarrow \left( -1\right)
^{\epsilon _1\epsilon _2}\left| \epsilon _1\epsilon _2\right\rangle ,\;
\end{equation}
where $\epsilon _1,\epsilon _2\in \{0,1\}.$ The transformation (5) implies
that only when the control qubit (the first qubit) is in the state $\left|
1\right\rangle ,$ a phase flip, i.e, a change from the sign $``+"$ to the
sign $``-",$ happens to the state $\left| 1\right\rangle $ of the target
qubit (the second qubit).

The cavity mode is initially in the vacuum state $\left| 0\right\rangle _c.$
The procedure for realizing a two-qubit CP gate is listed as follows:

Step (i): Apply a pulse (with a frequency $\omega =\omega _{31},$ a phase $%
\phi =-\frac \pi 2,$ and a duration $t_{1,a}=\frac \pi {2\Omega }$) to qubit
$1$ [Fig.~2(a)], to transform the state $\left| 1\right\rangle _1$ to $%
\left| 3\right\rangle _1$ as described by equation~(4). Wait for a time $%
t_{1,b}=\frac \pi {2g_1}$ to have the cavity mode resonantly interacting
with the $\left| 2\right\rangle \leftrightarrow \left| 3\right\rangle $
transition of qubit $1$ [Fig.~2(a$^{\prime }$)], such that the state $\left|
3\right\rangle _1\left| 0\right\rangle _c$ is transformed to $-i\left|
2\right\rangle _1\left| 1\right\rangle _c$ as described by equation~(2) while the
state $\left| 0\right\rangle _1\left| 0\right\rangle _c$ remains unchanged.
Then, apply a pulse (with a frequency $\omega =\omega _{21},$ a phase $\phi
=-\frac \pi 2,$ and a duration $t_{1,c}=\frac \pi {2\Omega }$) to qubit $1$
[Fig.~2(a$^{\prime \prime }$)], to transform the state $\left|
2\right\rangle _1$ to $-\left| 1\right\rangle _1$ as described by equation~(4).

\begin{figure}[tbp]
\centerline{\includegraphics[bb=23 222 583 670, width=9.6 cm, clip]{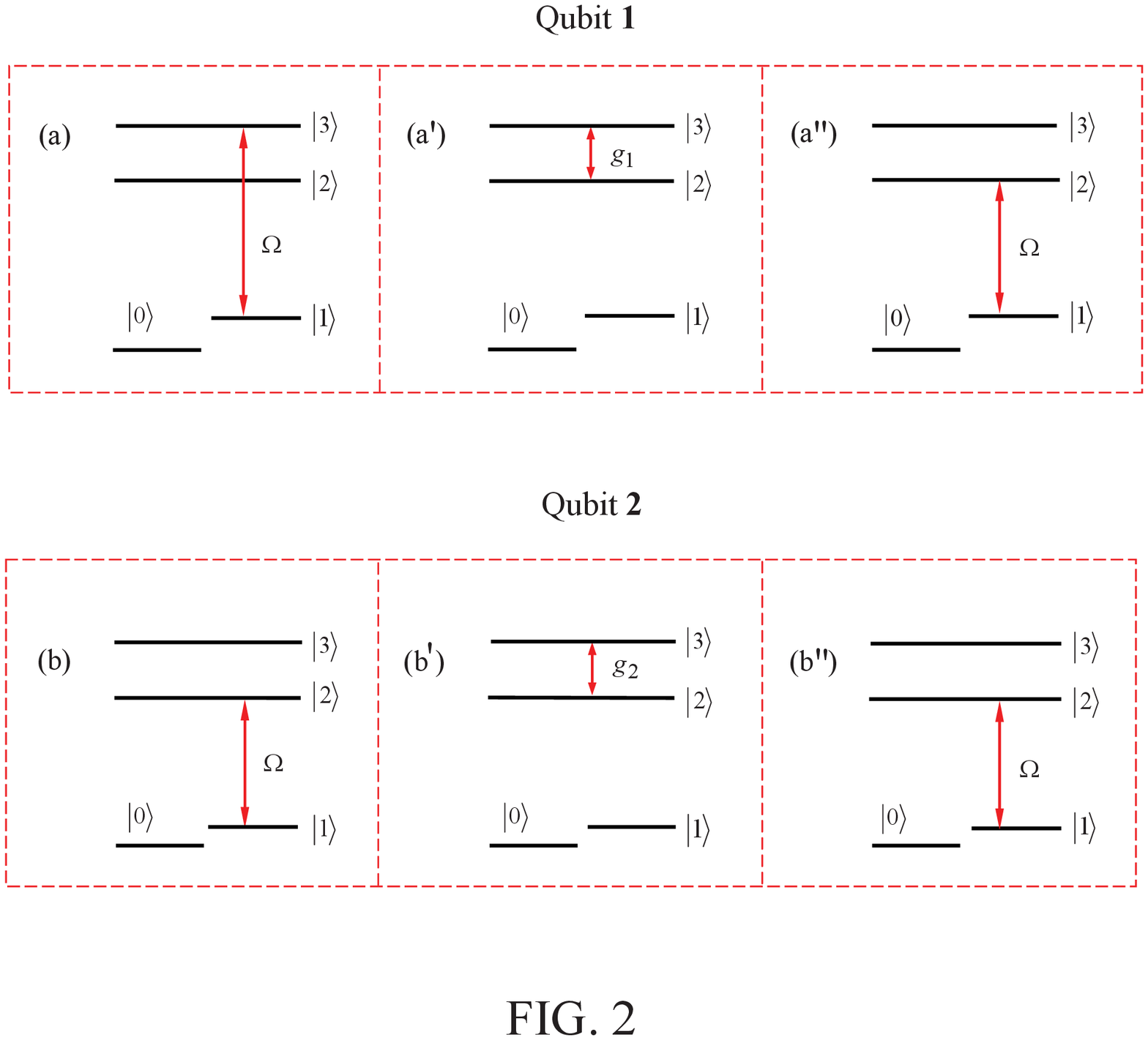}} %
\vspace*{-0.08in}
\caption{(Color online) Illustration of qubit 1 or qubit 2 interacting with
the cavity mode or the pulses during the gate operation. Here, the level
spacing between the top two levels $\left| 2\right\rangle $ and $\left|
3\right\rangle $ is set to be the same for each qubit; while the
level-spacing difference for any other two levels of each qubit is caused
due to nonuniformity in the qubit device parameters.}
\end{figure}

It can be checked that after the operation of this step, the following
transformation is obtained:

\begin{equation}
\begin{array}{c}
\left| 0\right\rangle _1\left| 0\right\rangle _c\otimes \left|
0\right\rangle _2 \\
\left| 0\right\rangle _1\left| 0\right\rangle _c\otimes \left|
1\right\rangle _2 \\
\left| 1\right\rangle _1\left| 0\right\rangle _c\otimes \left|
0\right\rangle _2 \\
\left| 1\right\rangle _1\left| 0\right\rangle _c\otimes \left|
1\right\rangle _2
\end{array}
\stackrel{\text{Step (i)}}{\longrightarrow }
\begin{array}{c}
\left| 0\right\rangle _1\left| 0\right\rangle _c\otimes \left|
0\right\rangle _2 \\
\left| 0\right\rangle _1\left| 0\right\rangle _c\otimes \left|
1\right\rangle _2 \\
i\left| 1\right\rangle _1\left| 1\right\rangle _c\otimes \left|
0\right\rangle _2 \\
i\left| 1\right\rangle _1\left| 1\right\rangle _c\otimes \left|
1\right\rangle _2
\end{array}
.
\end{equation}
The result (6) shows that after this step of operation, a photon is emitted
into the cavity, in the case when qubit 1 is initially in the state $\left|
1\right\rangle _1$ before the operation$.$

Step (ii): Apply a pulse (with a frequency $\omega =\omega _{21},$ a phase $%
\phi =-\frac \pi 2,$ and a duration $t_{2,a}=\frac \pi {2\Omega }$) to qubit
$2$ [Fig.~2(b)], to transform the state $\left| 1\right\rangle _2$ to $%
\left| 2\right\rangle _2$. Wait for a time $t_{2,b}=\frac \pi {g_2}$ to have
the cavity mode resonantly interacting with the $\left| 2\right\rangle
\leftrightarrow \left| 3\right\rangle $ transition of qubit $2$ [Fig.~2(b$%
^{\prime }$)], such that the state $\left| 2\right\rangle _2\left|
1\right\rangle _c$ changes to $-\left| 2\right\rangle _2\left|
1\right\rangle _c$ while the states $\left| 2\right\rangle _2\left|
0\right\rangle _c,$ $\left| 0\right\rangle _2\left| 0\right\rangle _c,$ and $%
\left| 0\right\rangle _2\left| 1\right\rangle _c$ remain unchanged. Then,
apply a pulse (with a frequency $\omega =\omega _{21},$ a phase $\phi =\frac
\pi 2,$ and a duration $t_{2,c}=\frac \pi {2\Omega }$) to qubit $2$ [Fig.~2(b%
$^{\prime \prime }$)], to transform the state $\left| 2\right\rangle _2$
back to $\left| 1\right\rangle _2.$

One can verify that after the operation of this step, the following
transformation is obtained:

\begin{equation}
\begin{array}{c}
\left| 0\right\rangle _1\otimes \left| 0\right\rangle _2\left|
0\right\rangle _c \\
\left| 0\right\rangle _1\otimes \left| 1\right\rangle _2\left|
0\right\rangle _c \\
i\left| 1\right\rangle _1\otimes \left| 0\right\rangle _2\left|
1\right\rangle _c \\
i\left| 1\right\rangle _1\otimes \left| 1\right\rangle _2\left|
1\right\rangle _c
\end{array}
\stackrel{\text{Step (ii)}}{\longrightarrow }
\begin{array}{c}
\left| 0\right\rangle _1\otimes \left| 0\right\rangle _2\left|
0\right\rangle _c \\
\left| 0\right\rangle _1\otimes \left| 1\right\rangle _2\left|
0\right\rangle _c \\
i\left| 1\right\rangle _1\otimes \left| 0\right\rangle _2\left|
1\right\rangle _c \\
-i\left| 1\right\rangle _1\otimes \left| 1\right\rangle _2\left|
1\right\rangle _c
\end{array}
.
\end{equation}

Step (iii): Apply a pulse (with a frequency $\omega =\omega _{21},$ a phase $%
\phi =-\frac \pi 2,$ and a duration $t_{1,c}=\frac \pi {2\Omega }$) to qubit
$1$ [Fig.~2(a$^{\prime \prime }$)], resulting in the transformation $\left|
1\right\rangle _1\rightarrow $ $\left| 2\right\rangle _1$. Wait for a time $%
t_{1,b}=\frac \pi {2g_1}$ to have the cavity mode resonantly interacting
with the $\left| 2\right\rangle \leftrightarrow \left| 3\right\rangle $
transition of qubit $1$ [Fig.~2(a$^{\prime }$)], such that $\left|
2\right\rangle _1\left| 1\right\rangle _c\rightarrow -i\left| 3\right\rangle
_1\left| 0\right\rangle _c$. Then, apply a pulse (with a frequency $\omega
=\omega _{31},$ a phase $\phi =\frac \pi 2,$ and a duration $t_{1,a}=\frac
\pi {2\Omega }$) to qubit $1$ [Fig.~2(a)], leading to $\left| 3\right\rangle
_1\rightarrow \left| 1\right\rangle _1.$

It can be checked that after the operation of this step, the following
transformation is obtained:

\begin{equation}
\begin{array}{c}
\left| 0\right\rangle _1\left| 0\right\rangle _c\otimes \left|
0\right\rangle _2 \\
\left| 0\right\rangle _1\left| 0\right\rangle _c\otimes \left|
1\right\rangle _2 \\
i\left| 1\right\rangle _1\left| 1\right\rangle _c\otimes \left|
0\right\rangle _2 \\
-i\left| 1\right\rangle _1\left| 1\right\rangle _c\otimes \left|
1\right\rangle _2
\end{array}
\stackrel{\text{Step (iii)}}{\longrightarrow }
\begin{array}{c}
\left| 0\right\rangle _1\left| 0\right\rangle _c\otimes \left|
0\right\rangle _2 \\
\left| 0\right\rangle _1\left| 0\right\rangle _c\otimes \left|
1\right\rangle _2 \\
\left| 1\right\rangle _1\left| 0\right\rangle _c\otimes \left|
0\right\rangle _2 \\
-\left| 1\right\rangle _1\left| 0\right\rangle _c\otimes \left|
1\right\rangle _2
\end{array}
\end{equation}

The equations~(6) and (8) show that during the operations of step (i) and step (iii)
on qubit 1 and the cavity, the states $\left| 0\right\rangle _2$ and $\left|
1\right\rangle _2$ of qubit 2 do not change. In addition, Eq.~(7) shows that
during the operation of step (ii) on qubit 2 and the cavity, the states $%
\left| 0\right\rangle _1$ and $\left| 1\right\rangle _1$ of qubit 1 remain
unchanged. This is because the cavity mode was initially assumed to be
resonant with the $\left| 2\right\rangle \leftrightarrow \left|
3\right\rangle $ transition but highly detuned (decoupled ) from the
transition between any other two levels of each qubit.

Based on the equations~(6-8), it can be found that the states of the whole system
after each step of the above operations are summarized in the following
equation:
\begin{equation}
\begin{array}{c}
\left| 0\right\rangle _1\left| 0\right\rangle _2\left| 0\right\rangle _c \\
\left| 0\right\rangle _1\left| 1\right\rangle _2\left| 0\right\rangle _c \\
\left| 1\right\rangle _1\left| 0\right\rangle _2\left| 0\right\rangle _c \\
\left| 1\right\rangle _1\left| 1\right\rangle _2\left| 0\right\rangle _c
\end{array}
\stackrel{\text{Step (i)}}{\longrightarrow }
\begin{array}{c}
\left| 0\right\rangle _1\left| 0\right\rangle _2\left| 0\right\rangle _c \\
\left| 0\right\rangle _1\left| 1\right\rangle _2\left| 0\right\rangle _c \\
i\left| 1\right\rangle _1\left| 0\right\rangle _2\left| 1\right\rangle _c \\
i\left| 1\right\rangle _1\left| 1\right\rangle _2\left| 1\right\rangle _c
\end{array}
\stackrel{\text{Step(ii)}}{\longrightarrow }
\begin{array}{c}
\left| 0\right\rangle _1\left| 0\right\rangle _2\left| 0\right\rangle _c \\
\left| 0\right\rangle _1\left| 1\right\rangle _2\left| 0\right\rangle _c \\
i\left| 1\right\rangle _1\left| 0\right\rangle _2\left| 1\right\rangle _c \\
-i\left| 1\right\rangle _1\left| 1\right\rangle _2\left| 1\right\rangle _c
\end{array}
\stackrel{\text{Step(iii)}}{\longrightarrow }
\begin{array}{c}
\left| 0\right\rangle _1\left| 0\right\rangle _2\left| 0\right\rangle _c \\
\left| 0\right\rangle _1\left| 1\right\rangle _2\left| 0\right\rangle _c \\
\left| 1\right\rangle _1\left| 0\right\rangle _2\left| 0\right\rangle _c \\
-\left| 1\right\rangle _1\left| 1\right\rangle _2\left| 0\right\rangle _c
\end{array}
.
\end{equation}
This result (9) demonstrates that a phase flip happens to the state $\left|
1\right\rangle _1\left| 1\right\rangle _2$ of the two qubits while the
cavity mode returns to its original vacuum state after the operations above.
Namely, a two-qubit CP gate described by equation~(5) is realized after the above
operations.

From the description above, it can be seen that the proposal presented here
does not require adiabatic passage (slow variation of the pulse Rabi
frequency), or a second-order large detuning $\delta =\Delta _c-\Delta $
during the entire operation. Here, $\Delta _c=\omega _{32}-\omega _c$ is the
first-order large detuning between the cavity frequency $\omega _c$ and the $%
\left| 2\right\rangle \leftrightarrow \left| 3\right\rangle $ transition
frequency $\omega _{32}$ of the qubits, while $\Delta $ $=\omega
_{32}-\omega $ is the first-order large detuning between the pulse frequency
$\omega $ and the $\left| 2\right\rangle \leftrightarrow \left|
3\right\rangle $ transition frequency $\omega _{32}$ of the qubits. In
addition, one can see that the present proposal does not require a
first-order large detuning $\Delta _c$ or $\Delta ,$ either. Note that only
\textit{resonant} qubit-cavity interaction and \textit{resonant} qubit-pulse
interaction are used in this proposal. In contrast, a second-order large
detuning or adiabatic passage was employed for the previous
proposals [22-27]. Thus, when compared with the previous
proposals [22-27], the gate
operation in this proposal can be performed faster by two orders of
magnitude.

In addition, it can be seen from the gate operation that this proposal does
not require adjustment of the level spacings of the qubits during the entire
operation, which however was needed by the previous
proposals [3,21]. Furthermore, since the qubit-cavity
coupling constants $g_1$ and $g_2$ are not required to be
identical, either nonuniformity in the qubit device
parameters (resulting in nonidentical qubit level spacings) or non-exact
placement of qubits in the cavity is allowed by this proposal.

Several points need to be addressed as follows:

(i) We note that for the gate implementation, four levels of each qubit are
necessary in order to have an irrelevant qubit (qubit 1 or qubit 2) to be
decoupled from the cavity mode during the gate operation.

(ii) The decay of the level $\left| 1\right\rangle $ of each qubit can be
suppressed by increasing the potential barrier between the two lowest levels
$\left| 0\right\rangle $ and $\left| 1\right\rangle $ [29].

(iii) For simplicity, we considered the identical Rabi frequency $\Omega $
for each pulse during the operations above. Note that this requirement is
unnecessary. The Rabi frequency for each pulse can be different and thus the
pulse durations for each step of operations above can be adjusted
accordingly.

(iv) During the gate operation, to have the effect of the qubit-cavity
resonant interaction during the pulse negligible, the pulse Rabi frequency $%
\Omega $ needs to be set such that $\Omega \gg g_1,g_2$.

\begin{figure}[tbp]
\centerline{\includegraphics[bb=51 234 556 677, width=8.2 cm, clip]{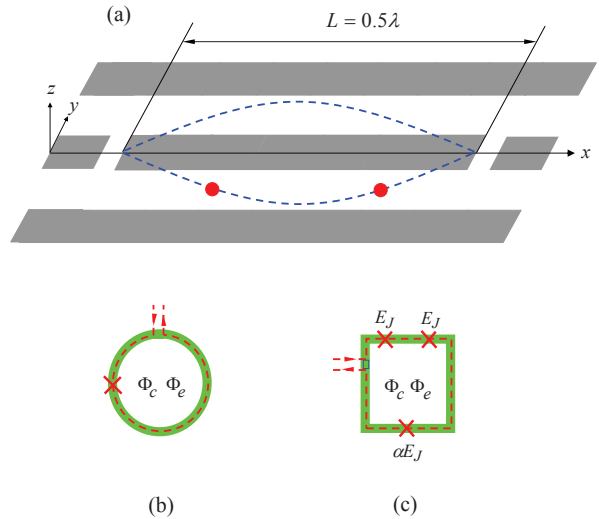}} %
\vspace*{-0.08in}
\caption{(Color online) (a) Setup for two superconducting flux qubits (red
dots) and a (grey) standing-wave one-dimensional coplanar waveguide
resonator. $\lambda $ is the wavelength of the resonator mode, and $L$ is
the length of the resonator. The two (blue) curved lines represent the
standing wave magnetic field in the $z$-direction. Each qubit (a red dot)
could be a radio-frequency superconducting quantum interference device (rf
SQUID) consisting of one Josephson junction enclosed by a superconducting
loop as depicted in (b), or a superconducting device with three Josephson
junctions enclosed by a superconducting loop as shown in (c). $E_J$ is the
Josephson junction energy ($0.6<\alpha <0.8$). The qubits are placed at
locations where the magnetic fields are the same to achieve an identical
coupling strength for each qubit. The superconducting loop of each qubit,
which is a large circle for (b) while a large square for (c), is located in
the plane of the resonator between the two lateral ground planes (i.e., the $%
x$-$y$ plane). For each qubit, the external magnetic flux $\Phi _c$ through
the superconducting loop for each qubit is created by the standing-wave
magnetic field threading the superconducting loop. A classical magnetic
pulse is applied to each qubit through an \textit{ac} flux $\Phi _e$
threading the qubit superconducting loop, which is created by an \textit{ac}
current loop (i.e., the red dashed-line loop) placed on the qubit loop. The
pulse frequency and intensity can be adjusted by changing the frequency and
intensity of the \textit{ac} loop current.}
\label{fig:5}
\end{figure}

\begin{center}
\textbf{IV. POSSIBLE EXPERIMENTAL IMPLEMENTATION}
\end{center}

As shown above, it can be
found that the total operation time $\tau $ is given by
\begin{eqnarray}
\tau =\pi /g_1+\pi /g_2+3\pi /\Omega .
\end{eqnarray}
The $\tau $ should be much shorter than the energy relaxation time and
dephasing time of the levels $\left| 2\right\rangle $ and $\left|
3\right\rangle $ (note that the level $\left| 1\right\rangle $ has a longer
decoherence time than both levels $\left| 2\right\rangle $ and $\left|
3\right\rangle $), such that decoherence, caused due to spontaneous decay
and dephasing process of the qubits, is negligible during the operation.
And, the $\tau $ needs to be much shorter than the lifetime of the cavity
photon, which is given by $\kappa ^{-1}=Q/2\pi \nu _c,$ such that the decay
of the cavity photon can be neglected during the operation. Here, $Q$ is the
(loaded) quality factor of the cavity and $\nu _c$ is the cavity field
frequency. To obtain these requirements, one can design the qubits (
solid-state qubits) to have sufficiently long decoherence time, and choose a
high-$Q$ cavity such that $\tau \ll \kappa ^{-1}.$

For the sake of definitiveness, let us consider the experimental possibility
using two identical superconducting flux qubits coupled to a one-dimensional
coplanar waveguide transmission line resonator [Fig.~3(a)]. For
superconducting qubits, the typical qubit transition frequency (which is the
transition frequency between the two lowest levels $\left| 0\right\rangle $
and $\left| 1\right\rangle $ in our present case) is between 5
and 10 GHz [4,5,11-19]. As an
example, let us consider two identical flux qubits
with the $\left| 0\right\rangle \leftrightarrow \left| 1\right\rangle $
transition frequency $\sim 5$ GHz, the $\left| 1\right\rangle
\leftrightarrow \left| 2\right\rangle $ transition frequency $\sim 10$
GHz, and the $\left| 2\right\rangle \leftrightarrow \left| 3\right\rangle $
transition frequency $\sim 3$ GHz. The qubits with these transition
frequencies may be available by designing the qubits with device parameters
chosen appropriately. Without loss of generality, assume $g_1/2\pi
\sim g_2/2\pi \sim 100$ MHz$,$ which is available in experiments (see, e.g.,
Refs.~[15,19,33]). By
choosing $\Omega /2\pi \sim 300$ MHz [33], we have $\tau
\sim 15$ ns. Note that the decoherence time of levels $\left| 2\right\rangle
$ and $\left| 3\right\rangle $ have not been measured in experiments, to the
best of our knowledge. However, we remark that a decoherence time of the
levels $\left| 2\right\rangle $ and $\left| 3\right\rangle ,$ which is
sufficiently longer than $15$ ns, may be available within the present
technique or in the near future due to the rapid development of superconducting
quantum circuits with long decoherence time [2]. In addition, for the qubits
considered here, the resonator frequency is $\sim 3$ GHz [15,33]. For a resonator
with this frequency and a quality factor $Q\sim 10^4$, we have $\kappa ^{-1}\sim 530$
ns, which is much longer than the operation time $\tau $ here. Note that
superconducting coplanar waveguide resonators with a (loaded) quality factor
$Q\sim 10^6$ have been experimentally demonstrated [34,35].

Finally, for superconducting qubits located in a microwave resonator, the
qubits can be well separated, because the dimension of a superconducting
qubit is 10 to 100 micrometers while the wavelength of the cavity mode for a
microwave superconducting resonator is 1 to a few centimeters [6,21]. As
long as the two qubits are well separated in space [Fig.~3(a)], the loop
current of one qubit affecting the other qubit and the direct coupling
between the two qubits are negligible, which can be reached by designing the
qubits and the resonator appropriately [6,21]. We should mention that
further investigation is needed for each particular experimental setup.
However, this requires a rather lengthy and complex analysis, which is
beyond the scope of this theoretical work.

\begin{center}
\textbf{V. CONCLUSION}
\end{center}

We have presented a way to fast realize a two-qubit
controlled-phase gate with four-level superconducting flux qubits in cavity
QED. As shown above, this proposal has the following advantages: (i) The
coupling constants of each qubit with the cavity are not required to be
identical, which makes neither identical qubits nor exact placement of
qubits in the cavity to be required by this proposal; (iii) No adjustment of
the level spacings of qubits during the entire operation is needed, thus
decoherence caused due to the adjustment of the level spacings is avoided in
this proposal; and (iv) Because only qubit-cavity resonant interaction and
qubit-pulse resonant interaction are used by this proposal, the gate can be
performed much faster when compared with the previous proposals.

\begin{center}
\textbf{ACKNOWLEDGMENTS}
\end{center}

This work was supported in part by the National Natural Science Foundation
of China under Grant No. 11074062, the Zhejiang Natural Science Foundation
under Grant No. Y6100098, the Open Fund from the SKLPS of ECNU, and the
funds from Hangzhou Normal University.

\end{document}